\documentclass{iau}
\usepackage{graphicx}

\title[Time Lag in Sgr~A$^\star$ IDV between the 90 and 102 GHz light curves]
{Time Lag in Sgr~A$^\star$ Intra-Day Variability between the light curves at 90 and 102 GHz}

\author[A.\ Miyazaki, M.\ Tsuboi \& T.\ Tsutsumi]
{Atsushi Miyazaki\,$^1$, 
Masato Tsuboi\,$^2$, \and  Takahiro Tsutsumi\,$^3$}

\affiliation{$^1$\,Korea Astronomy and Space Science Institute, \\ 
776 Daedeokdae-ro, Yuseong-gu, Daejeon 305-348, Republic of Korea \\ 
email: {\tt amiya@kasi.re.kr } \\[\affilskip]
$^2$\,Institute of Space and Astronautical Science, \\ 
Sagamihara, Kanagawa 252-5210, Japan \\
$^3$\,National Radio Astronomy Observatory, \\ 
1003 Lopezville Road, Socorro, NM 87801-0387, USA }

\pubyear{2012}
\volume{290}
\jname{Feeding Compact Objects: Accretion on All Scales}
\editors{C. M.\ Zhang, T.\ Belloni, M.\ Mendez \& S. N.\ Zhang, eds.}
\begin{document}

\maketitle

\begin{abstract}
We performed the observation of the flux densities of Sgr~A$^\star$ at 90 and 102~GHz on 6 April 2005 using the Nobeyama Millimeter Array in order to detect the time lag between these frequencies.  We constructed light curves covering a few hour with 1 min bin, and the Intra-Day Variability, which had a rising phase and intensity peak, of Sgr~A$^\star$ is clearly seen at both frequencies.
We calculated the \textit{z}-transformed discrete correlation function between the light curves of Sgr~A$^\star$ at 90 and 102~GHz.  The derived time lag of the flares at these frequencies was approximately zero, contrary to our expectations based on the previously reported time lag at lower frequencies.  If the radio flares of Sgr~A$^\star$ are explained by the expanding plasma model, the light curve at 90~GHz would be delayed with respect to the one at 102~GHz.  However, we could not find such a delay with statistical significance in our data.
\end{abstract}

\firstsection

\section{Introduction}

Sagittarius~A$^\star$ (Sgr~A$^\star$) is a compact source with emissions from radio to X-ray, and it is believed to be associated with the Galactic Center black hole (GCBH).  Sgr~A$^\star$ was observed at many wavelengths, and temporal flux variations were reported.  Time variability observation is a powerful tool to probe the structure and emission mechanisms.  
From mm to submm wave, Intra-Day Variability (IDV), flux variability with a timescale of a few hours, has been observed frequently (e.g., \cite[Miyazaki et al.\ 2004]{miya04}).  
The emitter of Sgr~A$^\star$ is in controversy but undoubtedly the emissions come from very near the GCBH.  One of possibilities that have been proposed as the radio emitter is expanding plasma ejected near GCBH. 
The VLA detected time lags of $\sim$20-30 min between IDVs at 22 and 43~GHz (\cite[Yusef-Zadeh et al.\ 2006, 2008]{YZ06,YZ08}).  These time lags are consistent with the model that the radio flux is emitted from adiabatically expanding plasma.  However, additional observations at other frequencies are needed to verify it. 

\section{Observation Data and Light Curves}

Aiming to detect a small time lag, we observed the flux densities of Sgr~A$^\star$ at 90 and 102~GHz on 6 April 2005 using the Nobeyama Millimeter Array (NMA) in Japan. 
The flux densities at 90 and 102~GHz were simultaneously obtained with 1~GHz bandwidth.  We alternately observed Sgr~A$^\star$ and NRAO530 (2.16~Jy in April 2005) as calibrator. 
The visibility data are restricted to the projected baseline larger than 25~$k\lambda$ in order to suppress the contamination from the extended components around Sgr~A$^\star$.  
Fig.\,\ref{fig1} shows the light curves of Sgr~A$^\star$ covering a few hour with 1~min bin at 90 and 102~GHz on 6 April 2005 using the NMA.  It is obviously the IDV had a rising phase and broad intensity peak.  The IDV timescale for a two-fold increase is estimated to be about 2 hour assuming that the increase has a constant gradient. 

\section{Cross Correlation Analysis}

\begin{figure}[htbp]
\begin{center}
  \begin{minipage}[t]{.47\textwidth}
    \begin{center}
      \includegraphics[width=5.7cm]{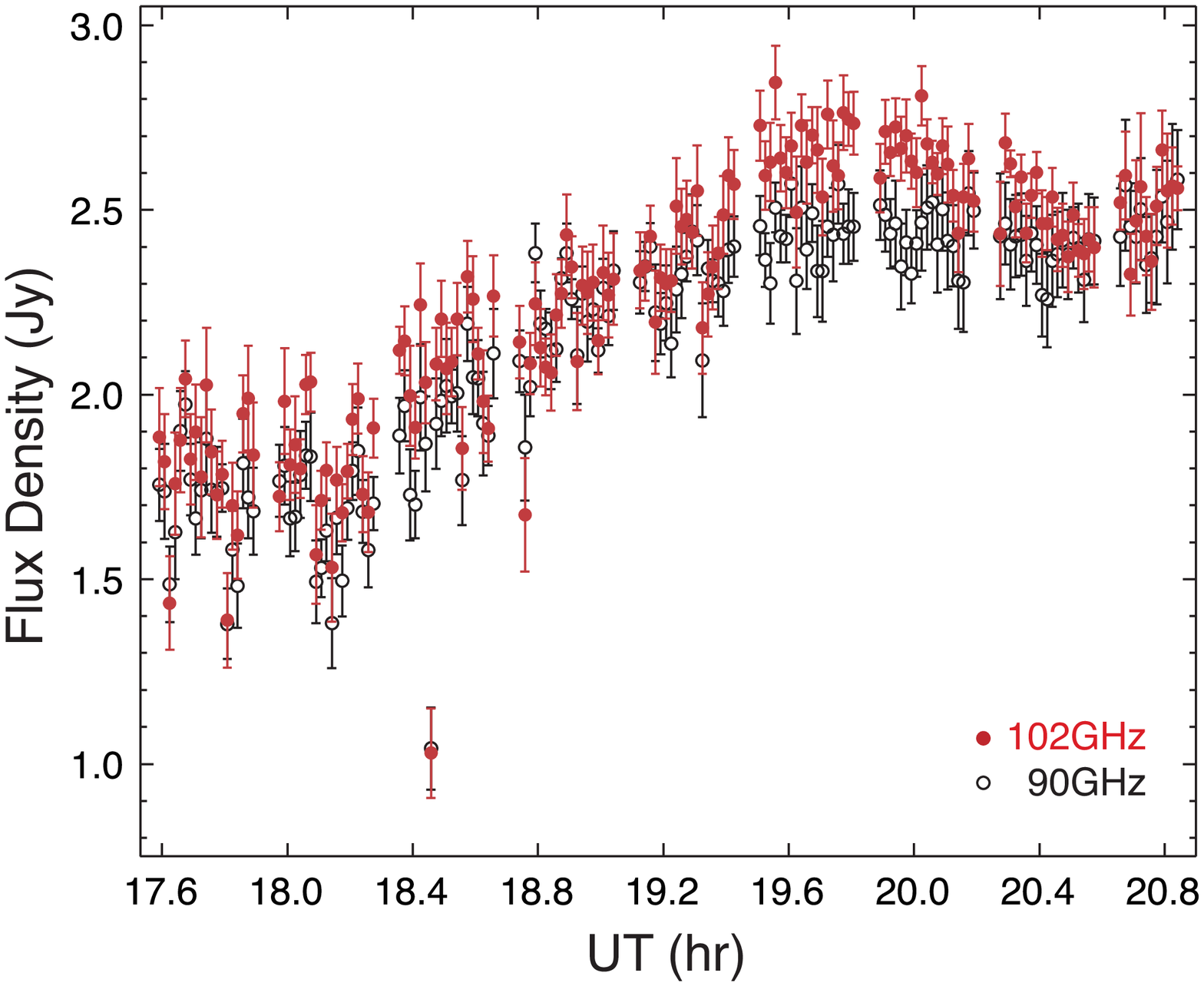}
      \caption{Sgr~A$^\star$ light curves using the NMA at 90 and 102~GHz on 6 April 2005.  Open and red filled circles indicate flux densities at 90 and 102~GHz, respectively.  The integration time of each data point is 1~min.  IDV, which shows a significant increase from 18.2~hr to 19.8~hr UT, was detected at both frequencies.  }
    \label{fig1}
    \end{center}
  \end{minipage}
  \hspace*{1.0mm}
  \begin{minipage}[t]{.47\textwidth}
    \begin{center}
      \includegraphics[width=6.0cm]{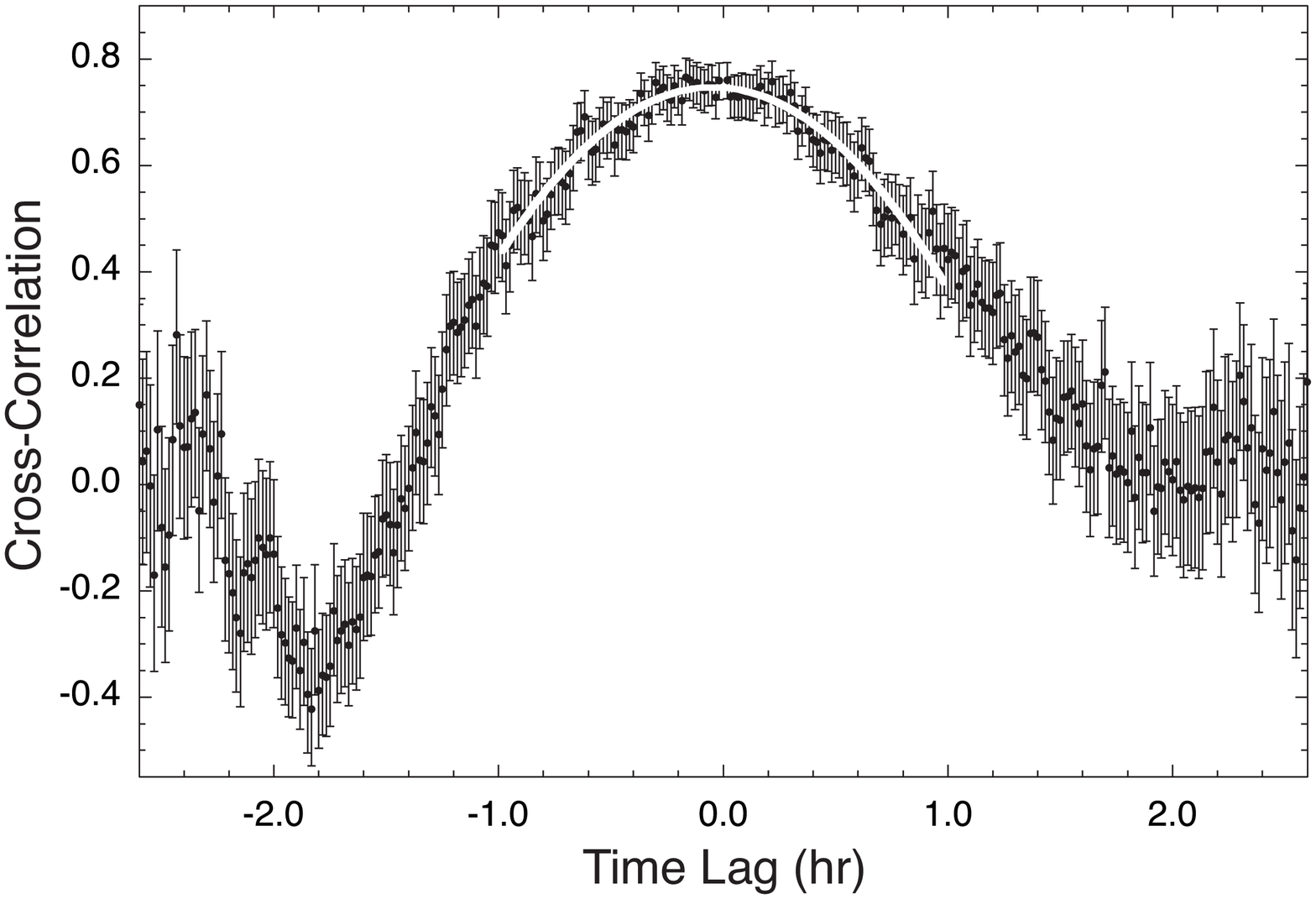}
      \caption{Cross-correlation function calculated with ZDCF of Sgr~A$^\star$ between 90 and 102~GHz.  There is a correlation function peak at time lag of approximate zero.  The peak time lag estimated by fitting the data to a quadratic function (thick curve) is $-2.56\pm0.92$~min. }
    \label{fig2}
    \end{center}
  \end{minipage}
\end{center}
\end{figure}

We calculated the \textit{z}-transformed discrete correlation function (ZDCF; \cite[Alexander 1997]{alex97}) to search for a time lag between the light curves of Sgr~A$^\star$ at 90 and 102~GHz on 6 April 2005 (see, fig.\,\ref{fig2}).  
The ZDCF appears to peak at a time lag of approximate zero.  The peak time lag estimated by fitting the data to a quadratic function in the range $\mid$time lag$\mid < 1$~hr is $-2.56\pm0.92$~min.  
The observed value indicates no time lag or the flux density at 102~GHz being marginally delayed relative to that at 90~GHz.  

The time lag of $\sim$20-30 min between the flux variability at 22 and 43~GHz reported by \cite[Yusef-Zadeh et al.\ (2006)]{YZ06} supports the expanding plasma model (\cite[van der Laan 1966]{vdL66}).  We estimated the time lag expected with the model based on \cite[Yusef-Zadeh et al.\ (2006, 2008)]{YZ06,YZ08}.  If the time lag $\Delta t_{43-22}$ is 25~min, the time lag $\Delta t_{102-90}$ is expected to be $+2.86$~min at $p=2$, the index of the relativistic electron energy spectrum.  The estimated time lag, $-2.56$~min, with NMA is sufficiently smaller than the expected value, $+2.86$~min. 
This result suggests that the expanding plasma blobs may be widely diverse.  Some blobs may be optically thin at 100~GHz even in the initial phase of the expansion.  Another possibility is that the major portion of the flux above 100~GHz does not originate from the blobs.
To reveal the origin of Sgr~A$^\star$ flare, more light curve samples are required.  

\vspace*{2mm}

\vspace*{-7mm}
 
\end{document}